\documentclass[osajnl,preprint,showpacs]{revtex4}
\usepackage{hyperref}
\usepackage{epsfig}

\begin{document}

%\draft

%\wideabs{
\title{Time and frequency domain solutions in an optical analogue of Grover's search algorithm}
\author{T.W. Hijmans, T.N. Huussen, and R.J.C. Spreeuw}
\affiliation{Van der Waals-Zeeman Instituut, Universiteit
van Amsterdam, Valckenierstraat 65, 1018 XE, Amsterdam, The
Netherlands.}

\email{hijmans@science.uva.nl}

\date{\today}
%\maketitle

\begin{abstract}
We present new results on an optical implementation of Grover's
quantum search algorithm. This extends previous work 
in which the transverse spatial mode of a light beam
oscillates between a broad initial input shape and a highly
localized spike, which reveals the position of the tagged item.
The spike reaches its maximum intensity after $\sim\sqrt N$ round
trips in a cavity equipped with two phase plates, where $N$ is the
ratio of the surface area of the original beam and the area of the
phase spot or tagged item. In our redesigned experiment the search
space is now two-dimensional. In the time domain we demonstrate
for the first time a multiple item search where the items appear
directly as bright spots on the images of a gated camera. In a
complementary experiment we investigate the searching cavity in
the frequency domain. The oscillatory nature of the search
algorithm can be seen as a splitting of cavity eigenmodes, each of
which concentrates up to 50\% of its power in the bright spot
corresponding to the solution.
\end{abstract}


%\pacs{03.67.Lx, 42.25.-p, 42.30.Kq}

\maketitle

\vspace{4mm}

\section{Introduction}
Of all algorithms proposed for implementation on quantum
computers, Grover's search algorithm \cite{Gro97,Gro97a} is
perhaps the most successful in terms of experimental
demonstrations. The algorithm has been successfully tested in a
variety of experiments, ranging from NMR to purely optical
realizations 
\cite{JonMosHan98,ChuGerKub98,KwiMitSch00,AhnWeiBuc00,DorLonAnd01,Bhattacharya02,Puentes04}.
The algorithm performs the following task: given a random list of
$N$ items, one of which is labeled, it seeks out the tagged item
in $O(\sqrt{N})$ computational steps. A classical search, based on
randomly testing items until the desired tagged one is found,
requires on average $\frac{1}{2} N$ tries. Hence for large $N$ the
Grover quantum algorithm leads to a quadratic speedup of the
search procedure compared to classical methods.

Whereas in general quantum interference and entanglement are
considered essential for quantum computation, it has been shown that 
entanglement is not essential for quantum searching and that it is possible to
implement Grover's algorithm relying only on interference \cite{Llo99}.
It has been demonstrated experimentally that one can construct a
purely optical method using classical light that closely follows
Grover's original quantum algorithm \cite{Bhattacharya02}. 
The underlying mapping is sometimes called a ``unary'' representation. 
There remains one essential difference between the classical 
implementation and the true quantum algorithm: the resources 
(essentially the minimal size of the apparatus) needed for the unary 
version scale linearly with $N$ whereas the resources for the 
quantum version scale only as $\log N$. 
Indeed, with this proviso, the unary mapping allows the implementation 
of any quantum algorithm for small numbers of qubits, even in the 
presence of entanglement \cite{Spr98,Spr01}.

In this paper we experimentally investigate an optical, unary,
implementation of Grover's search algorithm that differs in one
essential aspect from previous versions. As usual, the output of
the device oscillates between a
standard input and the sought solution. The period of the
oscillation is approximately $(\pi/2) \sqrt{N}$. After one half
period the output coincides exactly with the sought solution.

The oscillatory nature of the output in Grover's algorithm closely
resembles a Rabi oscillation of a two level atom, the role of the
two levels being taken over by the states corresponding to the
standard input and the solution, respectively. The fact that effectively only
two states participate, suggests that it is possible to bring
the system into a steady state consisting of a superposition
of the input state and the sought solution.

In the present experiment we compare two versions of
implementing Grover's algorithm. The first version we use a pulse of light
and observe oscillation of the transverse mode between a broad standard
input shape and a narrow spike, which corresponds to the
solution of the search problem. In the second version we use a
single frequency light source and exploit the resonant properties
of a Fabry-Perot type cavity into which the Grover iterator has
been embedded, to construct the aforementioned superposition state
of input state and solution.

Our optical implementation is similar to the one that was
published previously \cite{Bhattacharya02}, but has been modified
in several essential ways. The original experiment has been
redesigned completely, now using a two-dimensional search space
instead of a one-dimensional one, and a gated camera to
selectively look at the entire beam profile after a chosen number
of cavity round trips. The size of the search space in the present
experiment ranges between about 400 to 1000 items, depending on
the conditions chosen, and allowed us to demonstrate for the first
time a multiple item search. The tagged items appear directly as
bright spots in the camera images. In a complementary experiment
we feed the cavity with monochromatic light and scan the cavity,
observing Fabry-Perot resonances that are modified by the search.

The remainder of this paper is organized as follows: In the next
section we describe the principle of the experiment and its
interpretation. In section \ref{expdetail} we present the experimental
details, and in sections \ref{pulsresults} and \ref{cwresults} we present
results for the pulsed and cw versions of the experiment, respectively.
Finally, we discuss these results in section \ref{conclusion}.

\section{Principle of the experiment}
\label{principle}

In its quantum implementation Grover's algorithm starts out with
an $n-$qubit string which is prepared in a uniform superposition
of all its $2^n\equiv N$ states. A quantum operation called
Grover's iterator is then applied repeatedly to the $n-$qubit
string. This operation does the following: first the amplitude of
a single tagged state, which is unknown at the outset, is
inverted. The second step is an inversion of all amplitudes about
their average value. By repeating the Grover iterator $\sim\sqrt
N$ times the state is gradually transformed from the uniform
superposition into purely the tagged state: we have found our
solution.

The quantum algorithm has the following classical, unary, analog.
The search space is defined as the cross sectional area of a
coherent light beam of effective radius $R_b$. Let us divide the
transverse mode profile of this beam into $N$ small sections. The 
complex field amplitudes of the transverse beam profile in these 
$N$ sections represent the state vector. We
tag one of these sections by sending the beam through a phase
plate, called the ``oracle'', that contains a small circular spot that
locally changes the phase of the beam by $\pi /2$. It is natural
to define the size of the search space as $N=(R_b/r_s)^2$ where
$r_s$ is the radius of the phase spot. The phase spot is placed in
an arbitrary transverse position within the beam profile. The task at hand is
to identify the location of this phase spot in as few steps as
possible. This is done by passing the beam repeatedly through
a setup that has the property to progressively transfer more and
more intensity to the position of the phase spot. As soon as all
light intensity is concentrated in the spot we have found our
solution.

\begin{figure}
  \centering
  \includegraphics[width=\columnwidth]{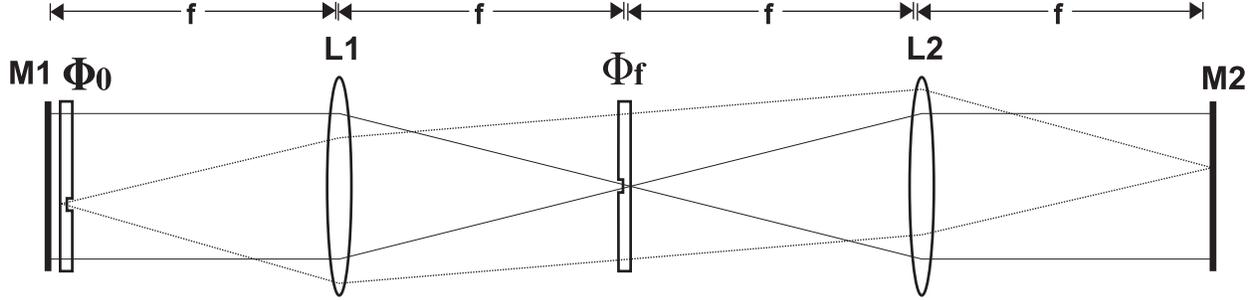}
  \caption{Optical cavity implementing the Grover iterator.
    M1: input mirror, M2: output mirror, L1, L2: lenses,
  $\Phi_{0}$: oracle phase-plate, $\Phi_{f}$: focus phase-plate  }
  \label{cavity}
\end{figure}

In Fig. \ref{cavity} we show the heart of the experimental setup.
It consists of a linear cavity with two mirrors with a
reflectivity of $90\%$. Inside the cavity there are two lenses of
focal length $f=0.34$ m, equal to $1/4$ of the cavity length. The
lenses are positioned at a relative distance of $2 f$ and
symmetric relative to the center of the cavity. The oracle phase
plate is positioned as close as possible to the input mirror of
the cavity and can be translated in two directions perpendicular
to the beam axis to ensure that the phase spot can be positioned
arbitrarily inside the beam profile. A second phase plate is
positioned at the common focus of the two lenses, in the center of
the cavity. Like the oracle, this phase plate contains a small
disk-shaped area in which the phase of the light wave is changed
by $\pi /2$. The size of the phase spots in both phase plates is
chosen equal to the waist of the light beam at the common focus of
the two lenses. The lenses and mirror are located such that the
transverse mode profile at the input mirror and oracle, and that
at the focus phase plate, are each others Fourier transform. It is
this property that makes this setup perform the equivalent of
Grover's algorithm. Each round trip both phase plates are passed
twice, leading to a local phase shift of $\pi$ per round trip. In
Table \ref{qccomp} we show the correspondence between the
operations of the quantum algorithm and its classical unary
equivalent.
\begin{table}
  \centering
  \begin{tabular*}{\columnwidth}{|lll  @{\extracolsep{\fill}} l|}
\hline
  & \textbf{Quantum}& & \textbf{Classical} \\
  & & &\\
  1. & Oracle & $ I - 2 | s \rangle \langle s | $  & $ \Phi_{0}^{2} = \exp[2 i \phi(x,y)] $  \\
  2. & Hadamard & $H^{\otimes n}$ &   $ \mathcal{F} $ \\
  3. &  & $ I - 2 | 0 \rangle \langle 0 | $ & $ \Phi_{f}^{2} = \exp[2 i \varphi(x,y)] $ \\
  4. & Hadamard & $H^{\otimes n}$ & $ \mathcal{F}^{-1} $ \\
\hline
\end{tabular*}
\caption{Comparison of quantum and classical operators in the
search protocol. Steps 2 through 4 in the quantum version are the
`inversion about average' (IAA) operation, $ \mathcal{F} $ denotes
Fourier transformation. $\phi$ and $\varphi$ denote the phase
shifts imparted by the oracle and focus phase plate, respectively.
The factor 2 inside the exponent appears because the setup is used
in double pass. }\label{qccomp}
\end{table}

In the notation of Table \ref{qccomp} we obtain the following
expression for the classical implementation of the Grover
iterator:
\begin{equation}\label{CGI}
  \mathbf{G} = \Phi_{0}^{2} \mathcal{F}^{-1} \Phi_{f}^{2} \mathcal{F}.
\end{equation}
Here the squares reflect the fact that the experiment uses a
double pass of each phase plate every round trip.

The normalized form of an intensity at position $\bf{r}$ which is
nonzero only in the circular region of one of the phase plates is
given as:
\begin{equation}
I_1(\mathbf{r})=\frac{1}{\pi r_s^2} \theta (r_s-|\mathbf{r}-\mathbf{r_i}|),
\label{spotintensity}
\end{equation}
where $\theta$ is the unit step function and $\bf{r_i}$ is the
location of the center of the phase spot. On the focus plate
$\mathbf{r}_i=0$, on the oracle it is the arbitrary position which we
wish to find.

In the experiment we use a Gaussian input beam. Unfortunately the
Fourier transform of a Gaussian does not exactly match the shape
of the circular focus phase spot. Therefore, in the theoretical
analysis of the experiment we will assume that the input beam has
the intensity profile of an Airy disk given in a form normalized
to unity when integrated over the radial coordinates by:
\begin{equation}
I_0(r)=\frac{k r_s}{4 \pi f^2 } [\frac{2J_1(\rho)}{\rho}]^2,
\label{Airy}
\end{equation}
where $\rho=k r_s r / f$, with $k$, $r$, and $f$, respectively
denoting the wave vector of the light, the distance from the
center in the input beam and the focal length of the lenses. The
number of channels is given as
\begin{equation}
N=4 f^2/k^2r_s^4. \label{numberofchannels}
\end{equation}
The amplitude corresponding to $I_0$ takes the role of the uniform
superposition of all possible $n$-bit strings used as input to
Grover's algorithm and the amplitude corresponding to $I_1$ takes
the role of the solution. Without loss of generality we can choose
these amplitudes $\psi_0(\textbf{r})$ and
$\psi_1(\bf{r},\bf{r_i})$ to be real. We can define the overlap of
these amplitude functions as:
\begin{equation}
\chi(\bf{r_i})=\int_{-\infty}^{\infty}\int_{-\infty}^{\infty}\psi_0(\bf{r})
\psi_1(\bf{r},\bf{r_i}) \rm{d}^2 \bf{r}. \label{overlap}
\end{equation}
The two functions $\psi_0$ and $\psi_1$ are almost but not quite
orthonormal. If $\bf{r_i}$ is chosen within the central disk of
the Airy pattern the overlap integral is of order $1/\sqrt N$. It
is useful to orthonormalize the two functions by retaining
$\psi_1$ and defining
\begin{equation}
\tilde{\psi}_0(\bf{r})= \frac{1}{\sqrt{1-\chi^2}}[\psi_0(\bf{r}) -
\chi(\bf{r_i})\psi_1(\bf{r},\bf{r_i})] \label{orthonormalize}
\end{equation}

We will now analyze the action of the setup of Fig. \ref{cavity}
for several cases using the above definitions. We will first
assume that the reflectivity of the mirrors is unity and that
additional losses are absent. When a light pulse performs a
complete round trip through the cavity each phase plate locally
imparts a $\pi/2$ phase change twice. One can show that a setup
were the phase plates impart a phase change of $\pi$ only once is
equivalent apart from trivial overall phase factors.
Experimentally this could be realized by using a ring cavity
rather than a linear one. We will adopt this assumption, which
also is implicit in Eq.~(\ref{CGI}), to slightly simplify the
analysis.

Using the two orthonormal amplitude functions $\tilde{\psi}_0$ and
$\psi_1$, one can easily show they transform in the following
manner after a single round trip of the cavity, up to an overall
phase factor depending on the cavity length:
\begin{eqnarray}
\tilde{\psi}_0 \rightarrow \tilde{\psi}_0 (1- 2\chi^2)+ 2\chi
\psi_1 \label{transform0}\\
\psi_1 \rightarrow \ \psi_1(1-2\chi^2) - 2\chi \tilde{\psi}_0,
\label{transform}
\end{eqnarray}
where we have suppressed the explicit dependence on the radial
coordinates. Eqs.~(\ref{transform0},\ref{transform}) are an approximation, valid up
to second order in $\chi$. It is derived by propagating the
amplitudes $\tilde{\psi}_0$ and $\psi_1$ through the setup,
imparting a phase shift of $\pi$ to the part of each function at the
position of the phase spots once each round trip.

If we start with all the power in the mode $\tilde{\psi}_0$, the
relative power in the two above modes at subsequent time $\tau$ is
expressed as the diagonal elements $\rho_{00}$ and $\rho_{11}$
which, to second order in $\chi$, take the form: $\rho_{00} =
\cos^2(2 \chi \tau)$ and $\rho_{11}=\sin^2(2\chi \tau)$. Here the
time $\tau$ is measured in units of cavity round trip time. This
time dependence is analogous to a Rabi oscillation in an undamped
two level atom, which reflects the well known fact that in
Grover's search problem the evolution is limited to a two level
subspace.

Note that the overlap $\chi$ actually depends on the relative
distance $\mathbf{r}_i$ of the oracle to the beam axis. If the oracle approaches
the origin the overlap becomes equal to $\chi=1/\sqrt{N}$. In this
case the amplitude of the solution is maximum for $\tau=\pi
\sqrt{N}/4$, corresponding to Grover's solution. If the oracle is
placed further outward, the search time increases slightly, to
become infinite for the oracle approaching a distance equal to the
first dark ring in the Airy disk.

The experiments described in the next section fall in two
categories. In the first experiment we proceed essentially as
just outlined, a small portion of the intensity of a
short pulse is admitted to the cavity through a mirror with high,
but finite reflectivity, in a beam profile closely resembling that
of Eq.~(\ref{Airy}). After each successive round trip the mode
evolves towards the sought solution. In a second category of
experiments we use cw light and look at the steady state solution
of the transverse mode in the cavity. We will assume that the
cavity is continually being fed light in mode $\psi_0$ through an
input mirror with amplitude transmission coefficient $t$. In the
cavity we assume that the transverse mode at the position of the
mirror, is given by $a_0 \tilde{\psi}_0 + a_1 \psi_1$. The losses
of the cavity are taken into account by assuming that they are
entirely due to the finite mirror reflectivity which leads to a
reduction of the amplitudes by a factor $\sqrt{1-t^2}$ for each of
the two mirrors of the cavity. The variable cavity length
introduces an overall phase factor $\exp [i\alpha]$ after each
round trip for both amplitudes. Experimentally this phase factor
can be varied either by sweeping the frequency of the light source
or by varying the cavity length. The condition of steady state
then leads, in analogy to Eq.~(\ref{transform}), to the following
equations:
\begin{eqnarray}
a_0=\{a_0(1-2\chi^2)- 2\chi a_1\}(1-t^2) \ \exp [i \alpha] + t \\
a_1=\{a_1(1-2\chi^2)+ 2\chi a_0\}(1-t^2) \ \exp [i \alpha],
\label{steadystate}
\end{eqnarray}
where the input power has been normalized to unity.

\begin{figure}
  \centering
  \includegraphics[width=65mm]{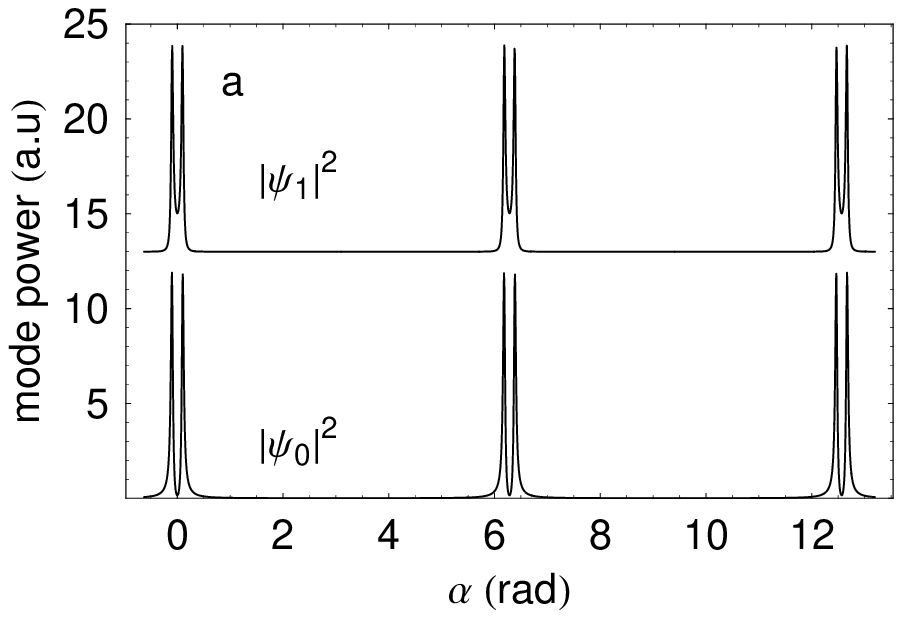}
  \includegraphics[width=65mm]{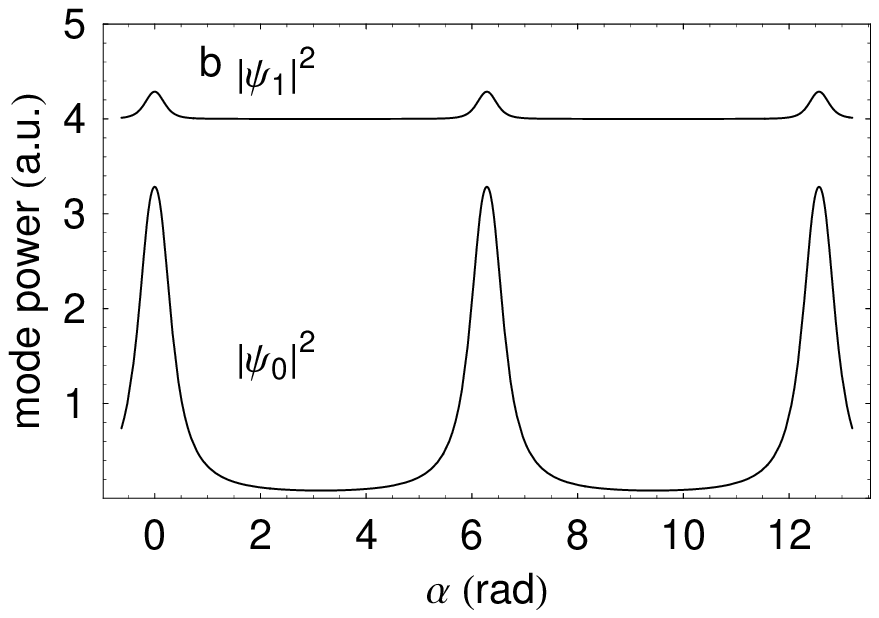}
  \caption{Calculation of the power corresponding to the eigenmodes  plotted versus the cavity
  phase shift angle $\alpha$,
   of a cavity with an oracle and a focus phase plate
  inserted, each imparting a local phase shift of $\pi$ each round trip. The lower trace in each figure
  corresponds to the mode $\tilde{\psi}_0$, the upper trace to $\psi_1$. The traces for $\psi_1$
  have been translated upwards for visual clarity. In both figures $\chi=0.05$. The a and b panels
  correspond to $t=0.15$ (a Finesse $F$ of $138$), and $t=0.5$ ($F=9$), respectively.}
  \label{cavitytheory}
\end{figure}

The solution of these equations is plotted in Fig.
\ref{cavitytheory} for $\chi=0.05$. This corresponds to a value of
$N=400$ a number roughly corresponding to the experimental
conditions below. In the figure we plot the calculated power
inside the cavity for both modes, as a function of the round-trip
phase shift. The result for both modes is similar to the normal
behavior for a resonant cavity, showing sharp maxima when the
effective cavity length is an integer number of half wavelengths.
However, on close inspection the resonance maxima are doublets
with a splitting that scales with $\chi$. The two lines of the
doublet are only resolved if the finesse $F$ of the cavity is of
the order of or bigger than $\pi/\chi$. For a two mirror
Fabry-Perot type cavity this finesse given by $F=\pi
\sqrt{R}/(1-R)$ where $R=1-t^2$.

What we see from Fig. \ref{cavitytheory} is that as soon as the
finesse of the cavity is sufficiently large, the transverse mode
is a doublet corresponding to the superpositions
$\frac{1}{\sqrt{2}}(\tilde{\psi}_0 \pm \psi_1)$. Almost half the power is
found in the solution $\psi_1$, for both components of the
doublet. In this doublet-resolved (``good cavity'') limit, the 
intensity of the solution, on resonance, is a factor $N$ higher than
in the background, i.e. in other positions. 

At this point it is instructive to consider the resource requirements. 
We could imagine setting the cavity on one of the doublet resonances and 
injecting a very low intensity, with on average a single photon inside the 
cavity. Since $F$ is essentially the number of roundtrips inside the cavity, and 
$F>\pi/\chi \sim\sqrt{N}$, this single photon will be detected after a time $O(\sqrt{N})$. 
As this photon has 50\% probability to be detected in the solution, only a small, 
fixed number of repetitions is needed. The usual $\sqrt{N}$ quantum search efficiency
thus appears here as a temporal resource requirement. 

In the opposite limit $F\ll\sqrt{N}$, where the doublet is unresolved, 
the quadratic speedup is lost. One can easily show that the contrast between
the solution and the background becomes $\sim F^2\ll N$. Note however that 
for a realistic finesse the contrast can still be substantially larger than unity. 
The setup can be thus be seen as a phase contrast microscope with a contrast
enhanced by several orders of magnitude. This may have applications beyond 
quantum information processing.

\section{Experimental details}
\label{expdetail}

The input light to the cavity comes from a 656 nm diode laser. The
laser in the pulsed experiment emits sub-ns pulses with a repetition rate of 40~MHz, in the
cw experiment we use a grating stabilized laser to obtain single
frequency cw light. The light is coupled into an optical fiber to
obtain a Gaussian TEM$_{00}$ beam profile. To implement the phase
shift of $\pi/2$, the phase plates consist of a fused silica substrate
with a dimple that is 359 nm deep (note that a bump
instead of a dimple would work as well). Phase plates of different
diameters ranging from $63$ to $127 \mu$m were used. From
Eq.~(\ref{numberofchannels}) we find that this corresponds to a
number of channels ranging from 300 to well over 1000. In most
experiments the effective channel number was about 400. The input
beam has a Gaussian shape which approximately matches the central
lobe of the Airy disc of the ideal case discussed in the previous
section. For phase spots of $100 \mu$m diameter the half width of
this Gaussian beam is about 2.5 mm.  The exact beam diameter was
optimized empirically in the experiment depending on which phase
plates were used.

The phase plates were manufactured using photolithography and
plasma etching. For technical reasons we decided to make holes
instead of bumps. The measured depth of the etched surfaces is
about 344 nm with a non-uniformity of at most 14 nm. The
corresponding phase shift is $1.51 \pm 0.06$ rad, which deviates
less than 8\% from the ideal $\pi/2$ value. In order to minimize
optical losses a multi-layer coating was applied to the phase
plates, reducing the reflectivity to less than 0.25\%.

\begin{figure}
  \centering
  \includegraphics[width=\columnwidth]{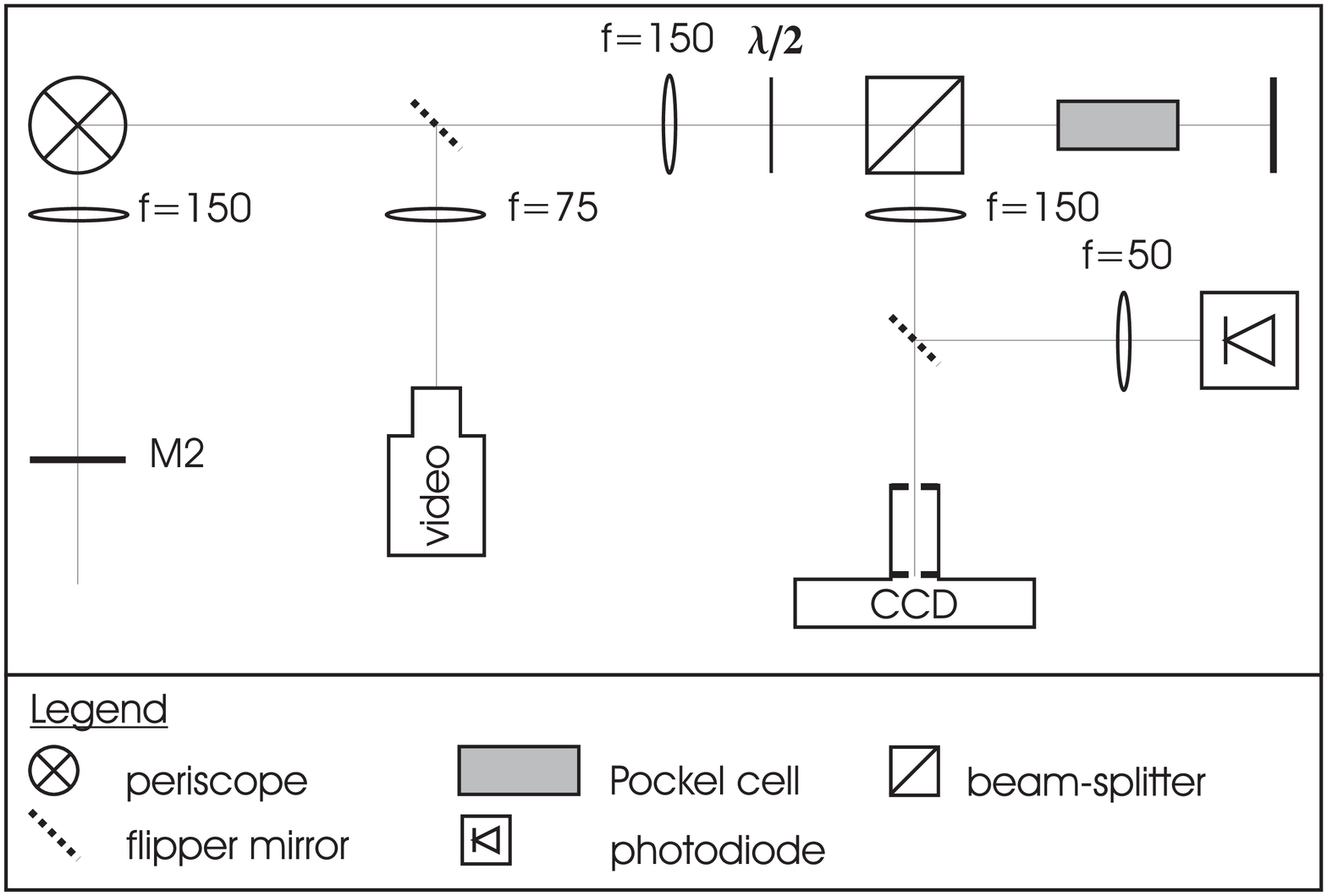}
  \caption{Setup for imaging. The image at mirror M2 is relayed and 
  imaged onto the CCD camera. 
  The half-wave plate rotates the polarization
  such that all light passes through the polarizing beam-splitter to the
  Pockels-cell. We use the Pockels-cell
  in double-pass configuration to obtain an effective retardation of $\lambda/2$, such
  that the retro-reflection is completely reflected by the beam-splitter. The
  last lens images the search result on the CCD-chip, or the photodiode. 
  The light path can be diverted to a video camera for real time
  monitoring of the image on the output mirror, used in the CW experiment. 
  In these experiments the video camera is replaced, after alignment, by
  a photo diode behind a $150 \mu$m aperture.}\label{imagesetup}
\end{figure}

In the 1D version of the pulsed experiment Bhattacharya et al.
\cite{Bhattacharya02} used a single photodiode for detection of
the output signal. The aim of our setup is to obtain a full 2D
picture of the search result. A CCD camera is suitable for this
purpose, but it is too slow to distinguish the pulses coming out
of the cavity, spaced in time by about 10 ns. Therefore we
implemented a setup for fast optical gating using a Pockels cell.
The layout of the imaging part of the setup is shown in Fig.
\ref{imagesetup}. The fast gate allows us to select a single pulse
from the pulse train that leaves the cavity. Each pulse
corresponds to $(i-\frac{1}{2})$ iterations ($i=1,2,3,\ldots$) 
of the search algorithm.  We are able to select
$i$-th pulse in repetitive measurements, such that we can
integrate the image on the camera. The high voltage supply to the
Pockels cell allows for a maximum repetition rate of 250 Hz and an
exposure time of about one minute yields good readout statistics.

In the cw experiments we used a fixed laser frequency and a galvo
driven rotating Brewster plate in the cavity to change its
effective length to obtain the spectra. The image at the output
mirror of the cavity was imaged onto a video camera to obtain a
qualitative impression of the spatial mode. For quantitative
analysis the camera was replaced by a photodiode with a $150 \mu$m
aperture in front. This photodiode and aperture were mounted on an
x-y translation stage allowing the transverse intensity profile to
be scanned. In the cw experiments the effective cavity length was
scanned over several wavelengths, using the Brewster plate. The
typical scan time was a few hunderd ms.

\section{Results of the time domain experiments}
\label{pulsresults}

\begin{figure}
  \centering
 \includegraphics[width= 60mm]{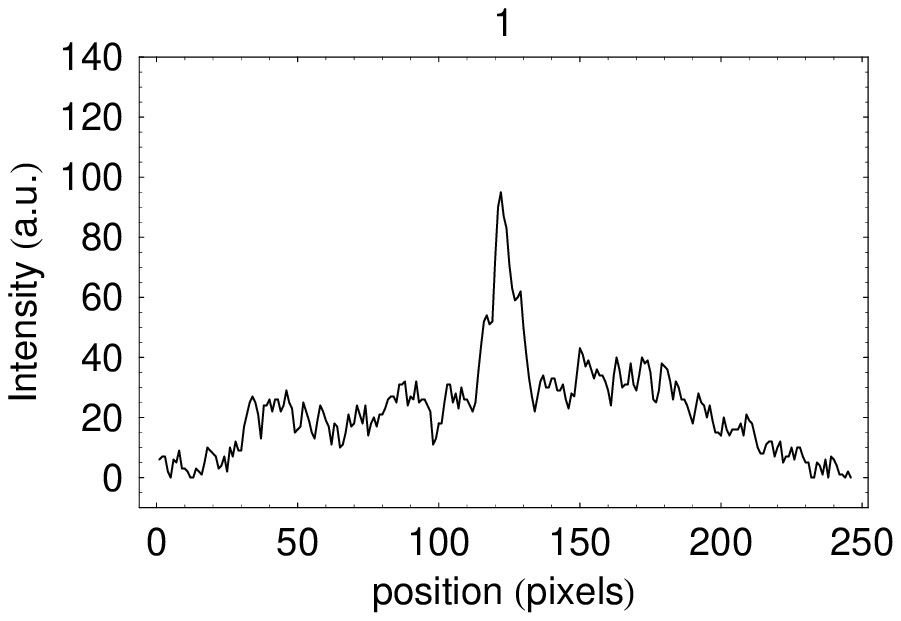}
  \includegraphics[width= 60mm]{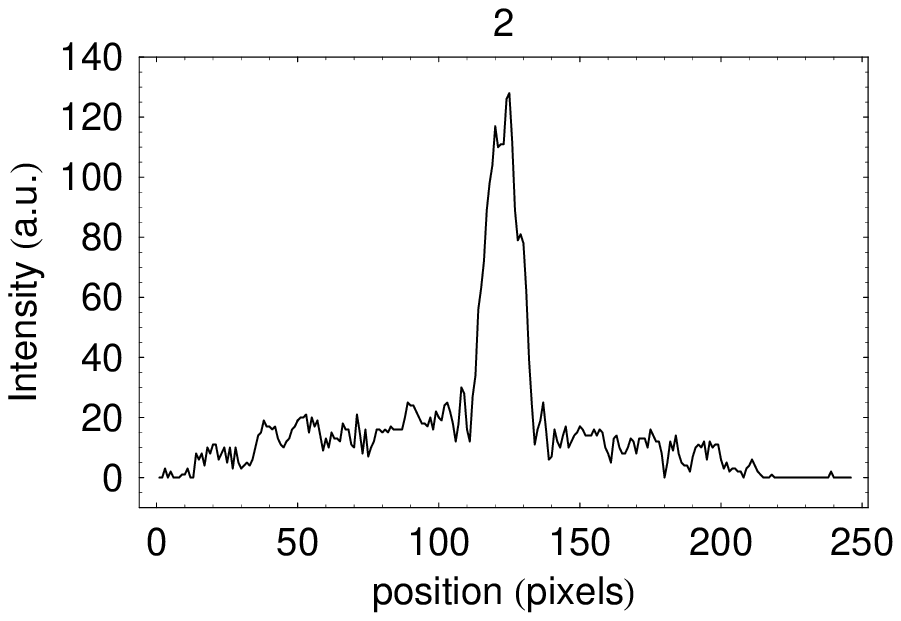}
   \includegraphics[width= 60mm]{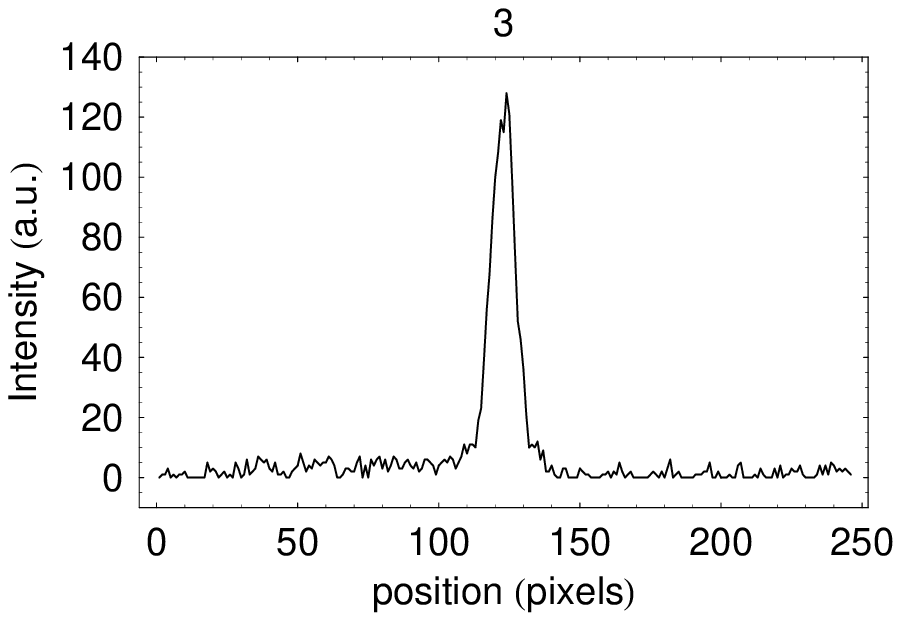}
    \includegraphics[width= 60mm]{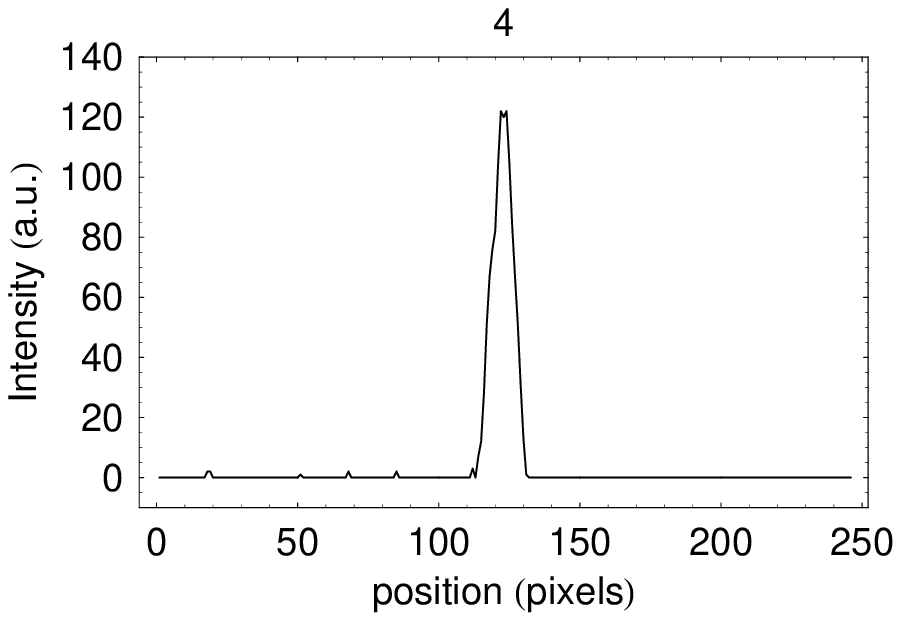}
% ImageAnalysis3.nb
  \caption{Cross section through the CCD image after a singe pass
  ($\frac{1}{2}$ round trip) and $1 \frac{1}{2}$,$2 \frac{1}{2}$, and $3 \frac{1}{2}$
  roundtrips. The images have the same scale and have not been corrected for losses.}
  \label{s-item}
\end{figure}

In Fig. \ref{s-item} we show the result of the pulsed version of
the experiment using an oracle containing a 63 $\mu$m diameter dot
together with a 90 $\mu$m diameter focus spot. Theoretically the
solution would reach a maximum after about 20 iterations. In the
figure we show a horizontal slice of the image obtained on the
CCD, going through the bright spot. The four images are taken
after $\frac{1}{2}$, $1 \frac{1}{2}$, $2 \frac{1}{2}$, and $3
\frac{1}{2}$ round trips, respectively. The shift by half a round
trip occurs because the light enters the cavity from the left and
the output is imaged on the right mirror (M2) in Fig. \ref{cavity}. It
is clearly seen that the ratio of the intensity of the sharp
feature to that of the broad background increases significantly
with each iteration. In Fig. \ref{s-item} we have subtracted the
light-free background recorded on the CCD. After 4 passes
the remnants of the signal corresponding to the input field have dropped below
zero due to slow drifts relative to the background signal. The images all have the same scale.

The sharp feature, the solution, does not display the expected
linear increase with time. The reason is that the cavity suffers
from considerable losses due to the finite transmittance of the
mirrors, reflectivity losses at the surfaces of lenses and phase
plates, and diffraction losses at the edges of the phase spots,
which after all are necessarily imperfectly matched to the
Gaussian input beam. The losses can be measured experimentally but
we defer a quantitative discussion of this measurement to the next
section. Here we can say that the results of Fig. \ref{s-item}
show that the solution reaches its maximum amplitude already after
2 or three round trips which implies that the losses amount to 30
or perhaps even $50 \%$ per complete round trip.

\begin{figure}
  \centering
  \includegraphics[width=\columnwidth]{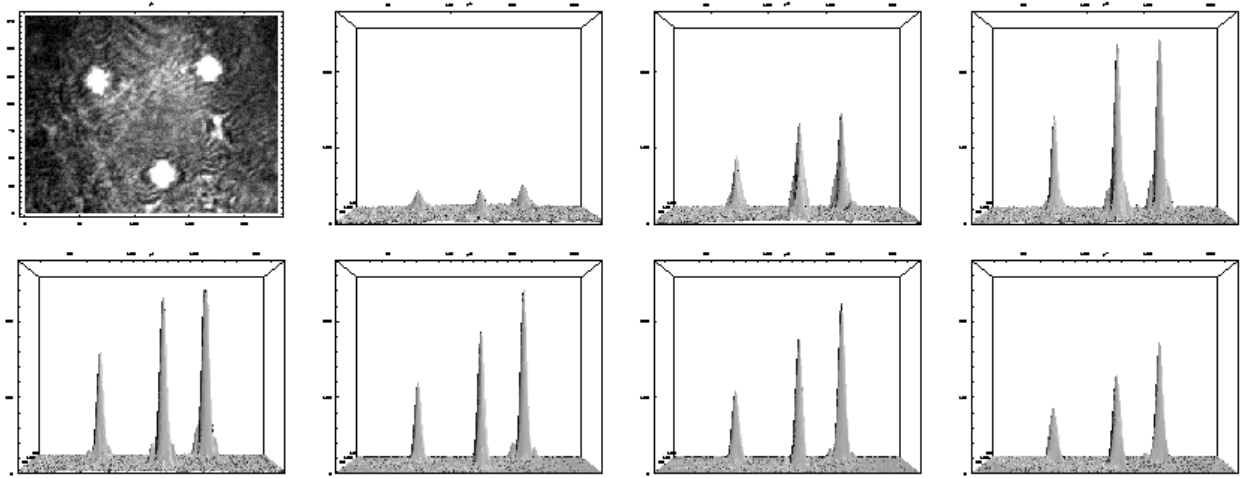}
  % ImageAnalysis8.nb
  % PROBLEM WITH FIGURE IN DVI FORMAT
  \caption{Multiple-item searching using the triple dot oracle.
  The upper left panel shows a grey scale image showing the
  three peaks in white. The remaining seven panels are 3-d plots of the peaks
  after $\frac{1}{2}$ through $6 \frac{1}{2}$ round trips. }
  \label{o1df4a}
\end{figure}

Before we turn to the cw results it is worth commenting briefly on a
version of Grover's algorithm in which $M > 1$ items are tagged. In
this case, starting from
the same state as in the one item case, a superposition of the $M$
solution states is reached after $O (\sqrt{N/M})$ iterations\cite{Gro97, Gro97a, NielsenChuang}. We
have investigated several examples of multi-item searching. A
representative example, where we use an oracle with three identical
phase spots of 100 $\mu$m diameter, is shown in Fig. \ref{o1df4a}.
The result is qualitatively similar to the one item case: the
contrast between the solutions and the background increases with
each iteration while the overall power drops rather quickly due to
the losses mentioned above.

\section{Results of the frequency domain experiments}
\label{cwresults}

\begin{figure}
  \centering
  \includegraphics[width=\columnwidth]{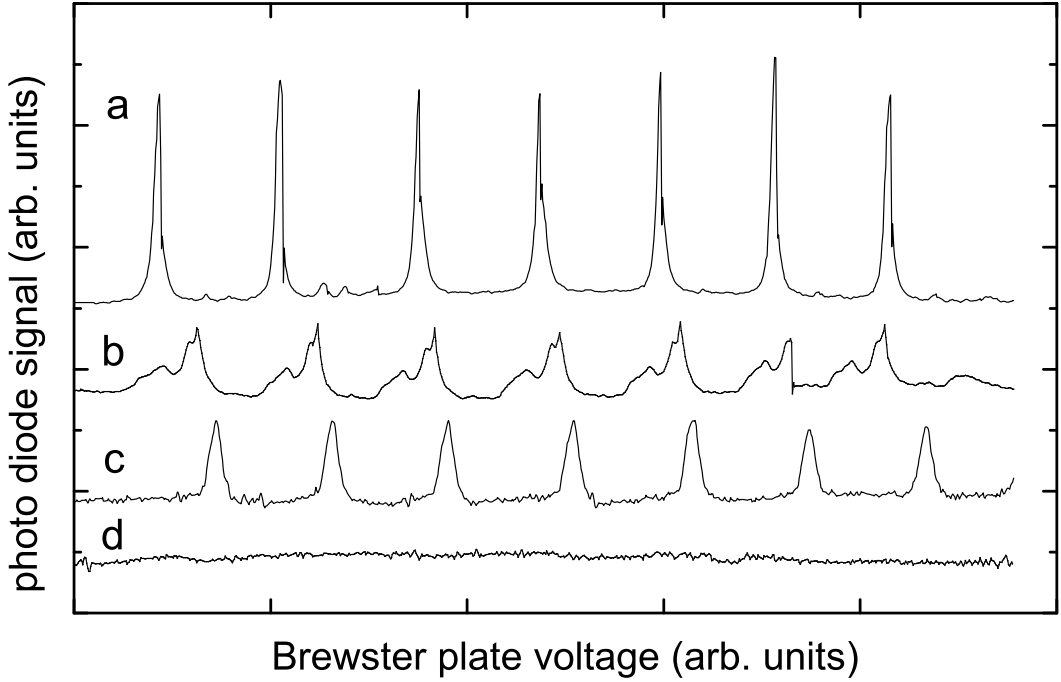}
  \caption{Photodiode signal taken at the image plane of the output mirror
  after a scanning cavity. Traces a and b are full
  images of the output power, taken with and without the focus phase plate. Traces c
  and d are taken with a $150 \mu$m aperture in front of the photo diode centered exactly on, and
  just beside, the position of the solution.}
  \label{fig:cwresults}
\end{figure}

As was discussed in section \ref{principle}, the setup shown in
Fig. \ref{cavity} is in essence a Fabry-Perot (F-P) resonator. By
using cw light and scanning the cavity length we can monitor its
resonance properties. First we measure the total power transmitted
at the output of the cavity with the phase plates positioned such
that both phase spots are completely outside the beam profile. The
result is the upper trace (a) shown in Fig. \ref{fig:cwresults}. The
characteristic peak structure of the F-P resonator is visible.
From this trace we find a finesse $F$ of about 20. If this were
entirely due to the mirror reflectivity we would infer 
$R=0.85$. In reality we use mirrors with $R=0.90$. The rather
low $F$ is consistent with this value of $R$ combined with the
losses of approximately $0.25\%$ per surface of the two lenses,
the phase plates and the Brewster plate inside the cavity, each
taken in double pass.

The second trace (b) in Fig. \ref{fig:cwresults} is the same as the
first, except that now the focus phase plate is positioned as
closely as possible to the waist of the beam in the focal plane.
Ideally this should give an identical spectrum, 
apart from an overall phase shift of $\pi$. In
reality the spectrum looks more messy. This is understandable in
view of the difficulty mentioned to mode match the waist in the
focal plane to the shape of the phase spot.

Finally, the lower two traces (c) and (d) in Fig. \ref{fig:cwresults}
constitute the actual cw experiment. Here the oracle phase spot is
placed inside the beam profile. Moreover, instead of measuring the
entire power transmitted through the output mirror of the cavity
we now image the plane of the output mirror onto the plane of the
$150 \mu$m aperture which is placed near the center of the beam
and just before the detector. Trace (c) showing the resonant
structure corresponds to the case where the position of the
aperture coincides with the position of the solution (the sharp
feature associated with the oracle phase spot) in the image on the
aperture plane. If the oracle phase spot is slightly moved so that
the spike falls just outside the region of the aperture, the
featureless trace (d) is measured. The signal to noise ratio of
the peaks in the spectrum is at about 20. No vestige of the peaks
can be seen just when the peak is moved outside the location of
the aperture indicating that the intensity of the solution is at
least 20 times higher than the background. The fact that we do not
see the predicted doublet structure in the cavity resonance is due
to the low finesse. From Fig. \ref{fig:cwresults}d we estimate the
value of $F$ to be about 9. This is consistent with about $50\%$
losses per round trip, roughly the same as in the pulsed experiment. As we see from Fig.
\ref{cavitytheory} for such a value of $F$ the power in the peak
is reduced by about a factor 20 as compared to the case of high
finesse. For $N=400$ this still gives a peak intensity a factor 20
above background, in excellent agreement with the observation.

Finally we should note that if we integrate the results of
Fig.~(\ref{fig:cwresults}) over frequency we still retain a strong
amplification of the solution. In other words, the scheme still
works with ``white light''. In this case the enhancement of the
solution is half of the value compared to the resonant
monochromatic case. The same is true if we integrate the time
domain results over time. This suggests practical applications
outside the direct field of quantum information. 

Seen from a different
perspective, the system essentially behaves as a phase contrast
microscope operated in a multi-pass fashion. The phase difference
imparted by the oracle is transformed into an amplitude contrast.
Grover's algorithm ensures that this contrast is enhanced by a
huge factor. This works for a single search item but also for a
multiple item oracle. One could well ask if this scheme could 
in principle be used to enhance the contrast of an image with an
arbitrary phase pattern. At first sight it
might seem that the usefulness of Grover's algorithm is rather
limited in this case. The reason is that the enhancement of the
amplitude of the search items upon iteration only occurs if the
phase shift imparted by the oracle items is the same as that of
the focus phase plate. On the other hand we may envisage using
this feature to our advantage. Grover's algorithm still works if
the total phase shift imparted by the oracle and the phase plate
are different from $\pi$ as long as they are equal. If we see our
oracle as a mask containing regions that give different phase
shifts, we can imagine using a focus phase plate with a variable
phase shift to selectively enhance the contrast of specific
regions that have just that phase shift. Such a variable-contrast
phase plate could possibly be constructed using a liquid crystal
device.

\section{Conclusion}
\label{conclusion}

We have demonstrated that Grover's algorithm can be implemented
using a classical optical arrangement relying only on interference
and without the need to invoke entanglement. The two dimensional
version of the experiment presented here provides a search space
of between 400 to 1000 items which would correspond to between 8
and 10 qubits in the quantum version. By exploiting the resonance
properties of the cavity and by using monochromatic cw light, we
have been able to demonstrate a version of the experiment that is
difficult, if not impossible, to implement in the quantum version
of the experiment. The transverse beam profile inside 
the cavity establishes a
steady state solution in this case that is a superposition of the
input state and the sought solution. The intensity in the sought
peak is more than an order of magnitude larger than that obtained
in a single pass. Our experiment may have practical applications
outside the direct field of quantum information, e.g. by providing 
new phase contrast imaging methods.

We acknowledge Chris Retif at the Amsterdam nanoCenter for assistance with producing
the phase plates used in the experiments. This work is part of the research program
of the Stichting voor Fundamenteel Onderzoek van de Materie (Foundation
for the Fundamental Research on Matter) and was made possible by
financial support from the Nederlandse Organisatie voor Wetenschappelijk
Onderzoek (Netherlands Organization for the Advancement of Research).


\begin{thebibliography}{99}
\expandafter\ifx\csname
natexlab\endcsname\relax\def\natexlab#1{#1}\fi
\expandafter\ifx\csname bibnamefont\endcsname\relax
  \def\bibnamefont#1{#1}\fi
\expandafter\ifx\csname bibfnamefont\endcsname\relax
  \def\bibfnamefont#1{#1}\fi
\expandafter\ifx\csname citenamefont\endcsname\relax
  \def\citenamefont#1{#1}\fi
\expandafter\ifx\csname url\endcsname\relax
  \def\url#1{\texttt{#1}}\fi
\expandafter\ifx\csname
urlprefix\endcsname\relax\def\urlprefix{URL }\fi
\providecommand{\bibinfo}[2]{#2}
\providecommand{\eprint}[2][]{\url{#2}}

\bibitem[{\citenamefont{Grover}(1997{\natexlab{a}})}]{Gro97}
\bibinfo{author}{\bibfnamefont{L.~K.} \bibnamefont{Grover}},
  \bibinfo{journal}{Phys.\ Rev.\ Lett.} \textbf{\bibinfo{volume}{79}},
  \bibinfo{pages}{325} (\bibinfo{year}{1997}{\natexlab{a}}).

\bibitem[{\citenamefont{Grover}(1997{\natexlab{b}})}]{Gro97a}
\bibinfo{author}{\bibfnamefont{L.~K.} \bibnamefont{Grover}},
  \bibinfo{journal}{Phys.\ Rev.\ Lett.} \textbf{\bibinfo{volume}{79}},
  \bibinfo{pages}{4709} (\bibinfo{year}{1997}{\natexlab{b}}).

\bibitem[{\citenamefont{Jones et~al.}(1998)\citenamefont{Jones, Mosca, and
  Hansen}}]{JonMosHan98}
\bibinfo{author}{\bibfnamefont{J.~A.} \bibnamefont{Jones}},
  \bibinfo{author}{\bibfnamefont{M.}~\bibnamefont{Mosca}}, \bibnamefont{and}
  \bibinfo{author}{\bibfnamefont{R.~H.} \bibnamefont{Hansen}},
  \bibinfo{journal}{Nature} \textbf{\bibinfo{volume}{393}},
  \bibinfo{pages}{344} (\bibinfo{year}{1998}).

\bibitem[{\citenamefont{Chuang et~al.}(1998)\citenamefont{Chuang, Gershenfeld,
  and Kubinec}}]{ChuGerKub98}
\bibinfo{author}{\bibfnamefont{I.~L.} \bibnamefont{Chuang}},
  \bibinfo{author}{\bibfnamefont{N.}~\bibnamefont{Gershenfeld}},
  \bibnamefont{and} \bibinfo{author}{\bibfnamefont{M.}~\bibnamefont{Kubinec}},
  \bibinfo{journal}{Phys.\ Rev.\ Lett.} \textbf{\bibinfo{volume}{80}},
  \bibinfo{pages}{3408} (\bibinfo{year}{1998}).

\bibitem[{\citenamefont{Kwiat et~al.}(2000)\citenamefont{Kwiat, Mitchell,
  Schwindt, and White}}]{KwiMitSch00}
\bibinfo{author}{\bibfnamefont{P.~G.} \bibnamefont{Kwiat}},
  \bibinfo{author}{\bibfnamefont{J.~R.} \bibnamefont{Mitchell}},
  \bibinfo{author}{\bibfnamefont{P.~D.~D.} \bibnamefont{Schwindt}},
  \bibnamefont{and} \bibinfo{author}{\bibfnamefont{A.~G.} \bibnamefont{White}},
  \bibinfo{journal}{J.\ Mod.\ Opt.} \textbf{\bibinfo{volume}{47}},
  \bibinfo{pages}{257} (\bibinfo{year}{2000}).

\bibitem[{\citenamefont{Ahn et~al.}(2000)\citenamefont{Ahn, Weinacht, and
  Bucksbaum}}]{AhnWeiBuc00}
\bibinfo{author}{\bibfnamefont{J.}~\bibnamefont{Ahn}},
  \bibinfo{author}{\bibfnamefont{T.~C.} \bibnamefont{Weinacht}},
  \bibnamefont{and} \bibinfo{author}{\bibfnamefont{P.~H.}
  \bibnamefont{Bucksbaum}}, \bibinfo{journal}{Science}
  \textbf{\bibinfo{volume}{287}}, \bibinfo{pages}{463} (\bibinfo{year}{2000}).

\bibitem[{\citenamefont{Dorrer et~al.}(2001)\citenamefont{Dorrer, Londero,
  Anderson, Wallentowitz, and A.Walmsley}}]{DorLonAnd01}
\bibinfo{author}{\bibfnamefont{C.}~\bibnamefont{Dorrer}},
  \bibinfo{author}{\bibfnamefont{P.}~\bibnamefont{Londero}},
  \bibinfo{author}{\bibfnamefont{M.}~\bibnamefont{Anderson}},
  \bibinfo{author}{\bibfnamefont{S.}~\bibnamefont{Wallentowitz}},
  \bibnamefont{and}
  \bibinfo{author}{\bibfnamefont{I.}~\bibnamefont{A.Walmsley}}, in
  \emph{\bibinfo{booktitle}{Proc. CLEO/QELS 2001}} (\bibinfo{publisher}{OSA},
  \bibinfo{address}{Washington DC}, \bibinfo{year}{2001}).

\bibitem{Bhattacharya02} N. Bhattacharya, H.B. van Linden van den
Heuvell, and R.J.C. Spreeuw, Phys. Rev. Lett. {\bf 88} 137901
(2002),

\bibitem{Puentes04} G. Puentes, C. La Mela, S. Ledesma, C. Iemmi,
J.P. Paz, and M. Saraceno, Phys. Rev. A {\bf 69} 042319 (2004).

\bibitem[{\citenamefont{Lloyd}(1999)}]{Llo99}
\bibinfo{author}{\bibfnamefont{S.}~\bibnamefont{Lloyd}},
  \bibinfo{journal}{Phys.\ Rev.\ A} \textbf{\bibinfo{volume}{61}},
  \bibinfo{pages}{010301(R)} (\bibinfo{year}{1999}).
  
\bibitem[{\citenamefont{Spreeuw}(1998)}]{Spr98}
\bibinfo{author}{\bibfnamefont{R.~J.~C.} \bibnamefont{Spreeuw}},
  \bibinfo{journal}{Found.\ Phys.} \textbf{\bibinfo{volume}{28}},
  \bibinfo{pages}{361} (\bibinfo{year}{1998}).
  
\bibitem[{\citenamefont{Spreeuw}(2001)}]{Spr01}
\bibinfo{author}{\bibfnamefont{R.~J.~C.} \bibnamefont{Spreeuw}},
  \bibinfo{journal}{Phys.\ Rev.\ A} \textbf{\bibinfo{volume}{63}},
  \bibinfo{pages}{062302} (\bibinfo{year}{2001}).

\bibitem{NielsenChuang} M.A. Nielsen and I.L. Chuang, Quantum
Computation and Quantum Information, Cambridge 2000.

\end{thebibliography}
\end{document}